\documentclass{aa}
\input epsfig.sty 

\begin{document}
   \thesaurus{03 (11.01.2; 11.09.1 IRAS~04575$-$7537; 11.17.2; 11.19.1; 13.25.2)}
   \title{ASCA detection of the {\rm Fe}\,K edge 
in the spectrum \\
of the Seyfert 2 galaxy IRAS~04575-7537:\\
a sign of a complex absorber}

   \author{C. Vignali
          \inst{1}
   \and A. Comastri
          \inst{2} 
   \and G.M. Stirpe
	  \inst{2}
   \and M. Cappi
          \inst{3,4}
   \and G.G.C. Palumbo
          \inst{1,3}
   \and M. Matsuoka
	  \inst{4}
   \and G. Malaguti
          \inst{3}
   \and L. Bassani
          \inst{3}
}

   \offprints{C. Vignali}
    \mail{l\_vignali@astbo3.bo.astro.it}
   \institute {Dipartimento di Astronomia, Universit\`a di Bologna, 
		Via Zamboni 33, I--40126 Bologna, Italy 
    \and Osservatorio Astronomico di Bologna,    
                 Via Zamboni 33, I--40126 Bologna, Italy   
    \and Istituto per le Tecnologie e Studio Radiazioni Extraterrestri 
	ITeSRE/CNR, via Gobetti 101, I--40129 Bologna, Italy
    \and The Institute of Physical and Chemical Research ({\em RIKEN}), 
		2-1, Hirosawa, Wako, Saitama, 351-01, Japan
    }

    \date{Received / Accepted}

   \maketitle

   \markboth{C. Vignali et al.: The discovery of the {\rm Fe}\,K edge in 
the ASCA Sey 2 spectrum of IRAS~04575$-$7537}{}
   \begin{abstract}
ASCA X-ray spectral analysis of the Seyfert 2 galaxy IRAS~04575$-$7537 
is presented. 
The main result is the presence of a significant iron edge at a rest energy of 
7.13$^{+0.21}_{-0.16}$ keV. 
%
%
The spectrum is flat ($\Gamma$ $\sim$ 1.5) with a substantial absorption 
N$_{\rm H}$ $\sim$ 10$^{22}$ cm$^{-2}$ and does not require any 
reflection component. There is also evidence of a narrow 
{\rm Fe}\,K$\alpha$ emission line, whose 
rest energy (E = 6.35$^{+0.08}_{-0.03}$ keV) and equivalent width 
(EW $\simeq$ 130$\pm{50}$ eV) suggest that the line originates from thick and cold matter. 
The intensity of 
the line and, in particular, the depth of the iron edge are too 
strong to be explained by simple transmission through the measured absorption 
column density. 
%
%
This strongly suggests that a model more complex than a single absorbed 
power law is needed. 
We propose an absorption model that we call ``leaky-total absorber'' 
which can explain the spectrum flatness, the iron emission line 
and the edge absorption feature and has also the advantage of having a 
straightforward physical interpretation in the framework of Unified models.  
In this model a thick absorber (N$_{\rm H}$ $\sim$ 10$^{23}$ cm$^{-2}$), 
possibly associated with broad line region (BLR) clouds, 
partially ($\sim$ 36\%) covers the source continuum with $\Gamma$ fixed 
to 1.9 as observed in Sey 1 galaxies. 
The escaping radiation is then absorbed by a 
column density N$_{\rm H}$ $\sim$ 10$^{22}$ cm$^{-2}$, which 
can be attributed to the torus, 
seen through its rim. 
%
%

 \keywords{X-rays: galaxies -- Galaxies: active -- Galaxies: Seyfert -- 
Galaxies: individual: IRAS~04575$-$7537 -- Galaxies: emission lines}
 \end{abstract}

%

\section{Introduction}

The main idea behind the Unification models for Active Galactic Nuclei 
(see Antonucci 1993 and references therein for a review) is that 
orientation plays a significant and decisive role in Seyfert classification. 
In terms of our current understanding of AGNs, the central engine is probably 
a massive black hole, surrounded by an accretion disk and fast-moving clouds, 
responsible for the production of strong gravitationally and Doppler-broadened 
lines, completely covered by a molecular torus. More distant uncovered 
clouds are, on the contrary, the origin of the narrow lines.
Studies of the X-ray spectra of Seyfert 2 galaxies (Awaki \& Koyama 1993) 
have revealed the presence of highly obscured nuclei 
with power law spectra and {\rm Fe}\,K$\alpha$ lines similar to Seyfert 1 objects 
providing further support to the popular Unification models for AGNs. 
If the orientation is such that the line of sight intercepts the 
torus (Krolik \& Begelman 1986), 
the Optical/UV radiation including the BLR emission as well as 
soft X-rays from the nucleus are blocked and the source, visible at these wavelengths 
in scattered light, 
is classified as a type 2. The scattering medium (Krolik \& Kallman 1987), 
whose physical state is unfortunately still poorly known, is likely to be a warm 
plasma located along the torus axis.\\
%
%
With respect to the hard X-ray emission, the spectral index 
distribution of Seyfert 2's should be the same as for Seyfert 1's. 
However this seems not to be the case, since discrepancies from the 
Unification scenario have been reported. In fact, 
the re-analysis of a sample of 15 Seyfert 2 galaxies observed 
by GINGA (Smith \& Done 1996) has indicated that the spectral slope of about 
half of them is significantly flatter ($\Gamma$ $\sim$ 1.5) than the mean value 
found for Seyfert 1 
($\Gamma$ $\sim$ 1.9-2.0, Nandra \& Pounds 1994; Gondek et al. 1996). 
At least 4 of these sources are incompatible with the steep Seyfert 1 slope, 
even with the addition to the primary spectrum of a strong reflection 
component. 
Moreover, Awaki (1991) has pointed out that a small subsample of IRAS selected 
Seyfert 2 galaxies shows no evidence of any obscuring medium 
($N_{\rm H}$ $<$ 10$^{22}$ cm$^{-2}$) and is well described by 
a rather steep continuum with no reflection component. 
%
%
On the basis of these results there seems to be growing evidence in favour 
of a more complex scheme for the X-ray emission of Seyfert 2 galaxies. 
To further investigate some of these issues we have observed with ASCA one of the 
galaxies in the Awaki (1991) subsample.\\
%
%
%
IRAS~04575$-$7537 is associated with a rather bright ($m_{\rm V}$ = 14.5) 
barred spiral galaxy ($z$ = 0.0184) and 
is characterized by a strong infrared emission, 
$F_{\rm FIR (40-120 \mu m)}$ $\simeq$ 4.2 $\times$ 10$^{-11}$ 
\mbox {erg cm$^{-2}$ s$^{-1}$}, corresponding to a 
$L_{\rm FIR}$ $\simeq$ 6.2 $\times$ 10$^{43}$ \mbox {erg s$^{-1}$} 
(Ward \& Morris 1984; Boller et al. 1992), which dominates over the 
optical emission (de Grijp et al. 1987, 1992). 
First detected in the X-ray band 
by the HEAO1 satellite (Wood et al. 1984), it was then observed 
by the GINGA/LAC (Awaki 1991; Smith \& Done 1996). 
It revealed a steep ($\Gamma$ $\simeq$ 2.1$\pm{0.2}$), 
mildly absorbed ($N_{\rm H}$ $\simeq$ 1.2$^{+0.7}_{-0.4}$ $\times$ 
10$^{22}$ cm$^{-2}$) continuum, 
plus an iron line (EW $\simeq$ 260$\pm{110}$ eV). 
It also appears as a bright X-ray source in 
the ROSAT All-Sky Survey (Moran et al. 1996), having a soft (0.1--2.4 keV) 
X-ray luminosity 
of about 10$^{43}$ \mbox {erg s$^{-1}$}, a typical value for a Seyfert 1 nucleus. 
Here we present an ASCA spectrum that allows a more detailed analysis of the 0.5--10 keV 
spectrum of this source.\\
Given that the optical classification of IRAS~04575$-$7537 
as a Seyfert 2 is based on a low signal-to-noise spectrum (Kirhakos \& Steiner 1990; 
de Grijp et al. 1992), we have also performed an optical observation, 
%
which is discussed in Sect. 2.  
Section 3 describes the ASCA observation and data reduction. Section 4 presents 
the X-ray timing and spectral analysis. 
The results are described in Sect. 5 and summarized in Sect. 6.\\
Throughout the paper a Friedmann cosmology with a 
Hubble constant $H_{0}$ = 50 \mbox {Km s$^{-1}$ Mpc$^{-1}$} 
and a deceleration parameter $q_{0}$ = 0 are assumed. \\

\section{Optical spectrum and classification}

\begin{figure}
\vspace{0cm}
\hspace{0cm}\epsfig{file=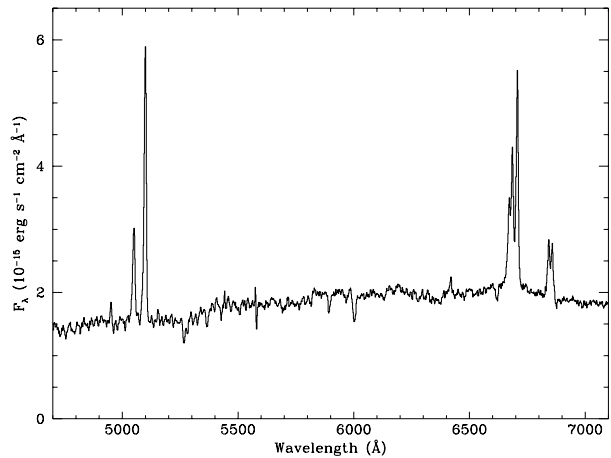, width=8.8cm, angle=-90, silent=}
\vspace{0cm}
\caption[]{The optical spectrum of IRAS~04575$-$7537, in the rest system of the observer 
(\rm z = 0.0184)}
\label{lab1}
\end{figure}

An optical spectrum of IRAS~04575$-$7537 was obtained in photometric conditions 
at the ESO 1.5m telescope on 1996 October 1, with a slit width of 2~arcsec, at a 
resolution of 4.6~\AA. The integration time was 30 minutes. Standard reduction 
and calibration techniques were used, and the resulting spectrum is shown in 
Fig.~\ref{lab1}. 

The spectrum presents several strong emission lines, none of which has a broad 
component, thus ruling out the Seyfert 1 galaxy type.  The fluxes of 
the main emission lines were obtained by fitting them with single Gaussians and 
integrating the latter.  The fluxes are listed in Table 1. As shown by 
Baldwin et al. (1981) and Veilleux \& Osterbrock (1987), ratios 
between fluxes of close lines can be used to determine whether a narrow emission 
line region is most likely to be photoionized by hot stars (as in starburst 
galaxies) or by an active nucleus (as in Seyfert 2 galaxies or in LINERS).  In 
the [O\,{\sc iii}]$\lambda$5007/H$\beta$ vs.\ [N\,{\sc ii}]$\lambda$6584/H$\alpha$ diagnostic 
diagram proposed by the above cited authors, the flux ratios of IRAS~04575$-$7537 (20 and 
1.5 respectively) place it well in the Seyfert 2 domain. 
The same occurs in the 
[O\,{\sc iii}]$\lambda$5007/H$\beta$ vs.\ [O\,{\sc i}]$\lambda$6300/H$\alpha$ and 
[O\,{\sc iii}]$\lambda$5007/H$\beta$ vs.\ [S\,{\sc ii}]$\lambda$6717 $+$ $\lambda$6731/H$\alpha$ 
diagrams.  The largest FWHM of the emission lines ($\sim$ 500~km~s$^{-1}$) is also 
typical for Seyfert 2 galaxies. 
We therefore confirm the Seyfert 2 classification and exclude a strong starburst 
component.

\begin{table}
\caption[]{Characteristics of the main optical lines}
\begin{flushleft}
\begin{tabular}{lll}
\hline\noalign{\smallskip}
Line & Total flux & EW\\
\smallskip
 & ($10^{-14}$ \mbox {erg s$^{-1}$ cm$^{-2}$}) & (Angstrom)\\
\hline\noalign{\smallskip}
H$\beta$ & 0.2 & 2\\
$\left[\ion{O}{iii}\right]$$\lambda$4959 & 1.7 & 10\\
$\left[\ion{O}{iii}\right]$$\lambda$5007 & 5.0 & 33\\
$\left[\ion{O}{i}\right]$$\lambda$6300 & 0.2 & 1\\
$\left[\ion{N}{ii}\right]$$\lambda$6548 & 1.6 & 7\\
H$\alpha$ & 2.5 & 12\\
$\left[\ion{N}{ii}\right]$$\lambda$6584 & 3.9 & 18\\
$\left[\ion{S}{ii}\right]$$\lambda$6717 & 1.0 & 6\\
$\left[\ion{S}{ii}\right]$$\lambda$6731 & 1.0 & 5\\
\noalign{\smallskip}
\hline
\end{tabular}
\end{flushleft}
\end{table}

\section{ASCA data reduction}
The observation reported here was carried out by the {\it Advanced Satellite for
Cosmology and Astrophysics} (ASCA) during the period 1996 November 4--5.
The satellite contains four sets of conical foil mirrors that focus X-rays into
four instruments covering the 0.4--10 keV energy band (Tanaka et al. 1994):
two solid-state spectrometers (SIS, Gendreau 1995), each consisting of 4 CCD chips yielding 
$\sim$ 2\% nominal energy resolution (FWHM) at 6 keV, and two gas imaging spectrometers
(GIS, Makishima et al. 1996) with $\sim$ 8\% energy resolution at 6 keV. 
The SIS data were obtained in 1-CCD
readout mode, whereby only 1 CCD chip is exposed on each SIS with the target at
the nominal pointing position. The FAINT mode data were converted into BRIGHT2
mode data (corrected for dark-frame error and for echo effect), in order to
minimize the effect of slight changes in the energy scale on the results.
The following selection criteria were applied to the data: 
the spacecraft had to be 
outside the South Atlantic Anomaly; the elevation angle above the Earth limb
greater than 5$\degr$; the bright Earth angle, i.e. the angle above the
Sun-illuminated Earth's limb, greater than 15$\degr$; the magnetic cut-off
rigidity (COR) greater than 6 GeV c$^{-1}$ for SIS and 7 GeV c$^{-1}$
for GIS. ``Hot'' and flickering pixels were removed from the SIS using an
algorithm that excludes pixels showing count rates outside the expected
Poissonian distribution estimated from neighbouring pixels; also 
rise-time rejection was applied to exclude particle events for the GIS data. 
SIS grades 0, 2, 3, 4 and 6 were selected
for the analysis. The introduction of grade 6 gives both an improvement in the source
photon number, expecially at energies above 5--6 keV (Weaver et al. 1997), and better
constraints to the spectral parameters. Moreover, the results are in agreement with 
the ones obtained with a more ``standard'' reduction without the introduction of the grade 6. 
The spectra were extracted from box regions of about 
8$\arcmin$ $\times$ 5$\arcmin$ 
for the SIS and from circular regions of radius $\sim$ 6$\arcmin$ 
for the GIS, both centered on the source. 
Background SIS spectra were taken from box source-free 
regions in the same chip which contained the source,
while for the GIS a circular background region was used from a position near the
source and at the same distance from the optical axis.

%
%
%
%
%
%
After removing a short period of unstable pointing prior to target acquisition, 
a total of $\sim$ 34 Ks/SIS and 35 Ks/GIS were available for the spectral 
analysis. 
%
%
The source count rates (in units of counts s$^{-1}$) are 
0.191$\pm$0.003 and 0.157$\pm$0.002 in the 0.4--10 keV band for SIS0 and SIS1, respectively, and 
0.151$\pm$0.002 and 0.153$\pm$0.003 in the 0.7--10 keV band for GIS2 and GIS3, respectively. 
Data preparation and spectral analysis were performed using 
version 1.3 of the {\sc XSELECT} package and version 9.01 of {\sc XSPEC} 
(Arnaud et al. 1996). 
%
%

\section{ASCA analysis and results}

\subsection{Timing analysis}

\begin{figure}
\vspace{0cm}
\hspace{0cm}\epsfig{file=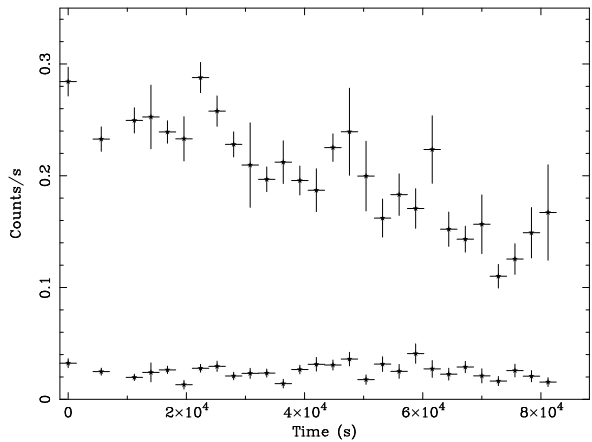, width=8.8cm, angle=-90, silent=}
\vspace{0cm}
\caption[ ]{IRAS~04575-7537 source+background light curve (top) 
and background light curve (bottom) in the 0.5--10 keV 
energy range. Bin interval is 2800 s}
\label{lab2}
\end{figure}

Figure~\ref{lab2} shows the 0.5--10 keV source+background (top) and background (bottom) 
light curves binned at 2800 s, 
which indicates that the background level (2.3 $\times$ 10$^{-2}$ counts s$^{-1}$) was 
stable during the ASCA pointing. 
The data are taken from SIS0, since no significant 
differences have been found among SIS and GIS. 
Flux variability with a doubling time of about 50\,000 s (thus constraining the dimensions 
of the emitting region to be of the order of 10$^{15}$ cm) 
and variations of the order 
of 35\% on shorter time scales of about 20\,000 s have been detected. 
The amplitude of the variability and the characteristic time-scales found for 
IRAS~04575$-$7537 are typical of Seyfert 1 galaxies (Mushotzky et al. 1993). 
It should be noted that no clear evidence of variability has been found 
in the soft (0.5--1.0 keV) light curve, 
whereas it is clearly present in the hard (2--10 keV) light curve. 
In order to look for spectral variability, 
a high- and a low-state spectrum were extracted, with about 0.2 counts s$^{-1}$ as the dividing line 
between the two states. 
Because of the rather poor statistics of the low-state spectrum, no 
conclusive information about spectral variability can be obtained. 
As a further check, the dividing line between the two states was choosen in such a 
way that the counting statistics was the same for both states. 
The derived high-state spectrum seems to be steeper than the low-state 
one, similar to what is observed in Seyfert 1 galaxies (Mushotzky et al. 1993). 
However, this result is only marginally significant and, therefore, 
it will not be considered any further. 
No significant flux variability has been detected between 
ASCA and GINGA observations on a time scale of 6 years. 
%

%
%

%

\subsection{Spectral analysis}

GIS and SIS spectra were binned with more than 20 counts per bin in 
order to apply the $\chi^{2}$ statistics. In the following, errors are 
given at 90\% of confidence for one interesting parameter 
($\Delta\chi^{2}$ = 2.71, Avni 1976). 
At first, a simple absorbed power law model was fitted to the data of each 
SIS/GIS detector. 
Since the results were all in agreement, within the errors, 
%
%
and the residuals from the fit were very similar, the two SIS 
and the two GIS spectra were added together. 
In the analysis described below the data were, therefore, 
fitted simultaneously using the same model, 
with the relative normalizations of SIS and GIS left free to vary. The best-fit parameters 
are given in Table 2. 

\begin{figure}
%
\vspace{0cm}
\hspace{0cm}\epsfig{file=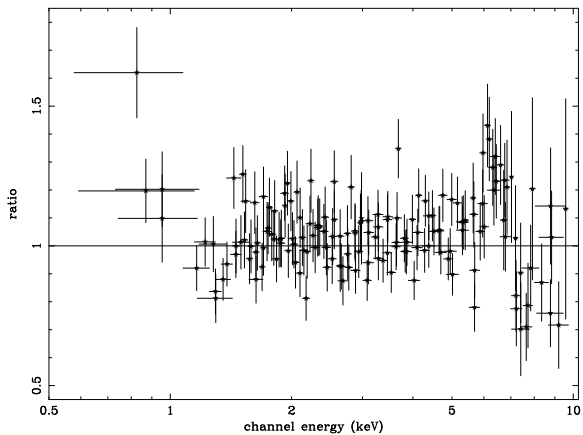, width=8.8cm, angle=-90, silent=}
\vspace{0cm}
\caption[ ]{Plot of the data/model ratio. The presence of an iron line + edge is suggested 
by the residuals in the power law plus absorption model. A soft component is 
also evident below $\sim$ 1 keV} 
\label{lab3}
\end{figure}

A simple power law continuum plus photoelectric absorption from cold material with 
cosmic abundances given by Anders \& Grevesse (1989) and 
cross sections derived from Balucinska-Church \& McCammon (1992) does not provide an adequate 
fit to the data. The plot of the data to model ratio (Fig.~\ref{lab3}) 
clearly reveals the presence of an iron line emission at about 6.4 keV plus 
an absorption edge at about 7 keV, as well as excess emission below 1 keV. 
Given the poor 
statistical quality of the data below 1 keV, the following will mainly focus 
on the X-ray properties of IRAS~04575$-$7537 at energies $>$ 1 keV. 
%

\begin{table*}
\caption[]{Results of spectral fits.\\
$\tau_{\rm edge}$ is the absorption depth of the iron edge, 
$\Re$ represents the reflection parameter, where $\Re$ = 1 for 
isotropic illumination}
\begin{flushleft}
\begin{tabular}{llccclccc}
\noalign{\smallskip}
\hline
\noalign{\smallskip}
$\Gamma$$_{\rm soft}$&$\Gamma$$_{\rm hard}$&N$_{\rm H}$
&$E_{\rm line}$&EW&$E_{\rm edge}$&$\tau_{\rm edge}$&$\Re$&$\chi^{2}/dof$\\
\noalign{\smallskip}
&&(10$^{22}$ cm$^{-2}$)&(keV)&(eV)&(keV)&&&\\
\noalign{\smallskip}
\hline \hline
\noalign{\smallskip}

\dots&1.47$^{+0.05}_{-0.04}$&1.03$\pm{0.06}$&\dots&\dots&\dots&
\dots&\dots&867/814\\

\dots&1.52$\pm{0.05}$&1.06$\pm{0.06}$&6.35$^{+0.07}_{-0.03}$&165$\pm{49}$&
\dots&\dots&\dots&837/812\\

\dots&1.43$\pm{0.06}$&0.99$^{+0.07}_{-0.06}$&6.35$^{+0.08}_{-0.03}$&
129$^{+46}_{-47}$&7.13$^{+0.21}_{-0.16}$&0.33$^{+0.14}_{-0.13}$&\dots&
819/810\\

%
%
%
\dots&1.48$^{+0.21}_{-0.10}$&1.01$^{+0.13}_{-0.07}$&6.35$^{+0.09}_{-0.04}$&
115$^{+62}_{-61}$&7.13$\pm{0.18}$&0.34$^{+0.14}_{-0.13}$&$<$ 3.24&819/809\\

\dots&1.90 fr.&1.24$^{+0.05}_{-0.04}$&6.35$^{+0.10}_{-0.08}$&72$^{+53}_{-46}$&
7.05$^{+0.19}_{-0.17}$&0.32$^{+0.14}_{-0.13}$&6.46$^{+1.05}_{-1.15}$&
830/810\\

1.49$^{+0.07}_{-0.06}$&$\Gamma_{\rm s}$ = $\Gamma_{\rm h}$&1.05$\pm{0.10}$
&6.35$^{+0.08}_{-0.03}$&142$^{+49}_{-50}$&7.14$^{+0.25}_{-0.17}$
&0.29$\pm{0.14}$&\dots&809/809\\
\noalign{\smallskip}
\hline
\end{tabular}
\end{flushleft}
\end{table*}

\begin{figure}
\vspace{0cm}
\hspace{0cm}\epsfig{file=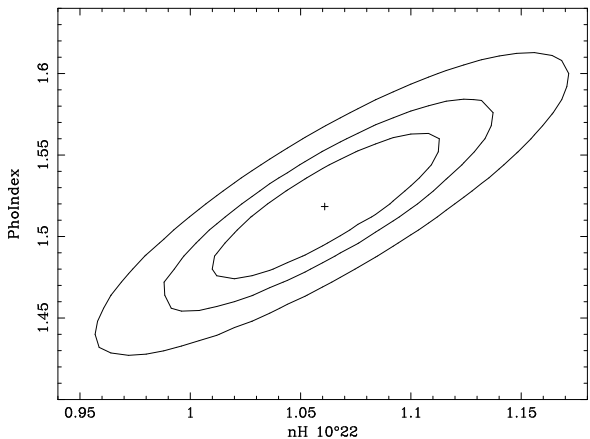, width=8.8cm, angle=-90, silent=}
\vspace{0cm}
\caption[ ]{Plot of the confidence contours for the spectral slope and 
absorption column density (second line of Table 2). 
The contours represent 68, 90 and 99 \% confidence intervals}
\label{lab4}
\end{figure}

Besides the spectral features at 6-7 keV, 
a single absorbed ($N_{\rm H}$ $\simeq$ 10$^{22}$ cm$^{-2}$) 
power law gives a flat slope, $\Gamma$ $\simeq$ 1.4-1.5 (Fig.~\ref{lab4}), which is significantly 
different from the GINGA slope $\Gamma$ $\simeq$ 2.1$\pm{0.2}$ 
(Smith \& Done 1996). 
The steeper GINGA spectrum can be, at least partially, 
explained in terms of contamination of the GINGA data by a nearby 
K\,{\sc iii} peculiar star, which is at  a distance of about 18$\arcmin$ 
in the GIS field of 
view from IRAS~04575$-$7537, has a 2--10 keV flux of $\sim$ 7 $\times$ 10$^{-12}$ 
\mbox {erg cm$^{-2}$ s$^{-1}$} (to be compared with the Seyfert 2 flux of 
1.2 $\times$ 10$^{-11}$ \mbox {erg cm$^{-2}$ s$^{-1}$}) 
and has a steep power law continuum with $\Gamma$ = 2.7$\pm{0.2}$. 
%
%

A net improvement in the spectral fit is achieved after 
the addition of a gaussian line at 6.4 keV 
($\Delta\chi^{2}$ = 30). If the gaussian width is kept 
narrow ($\sigma$ = 0 eV) and the central energy and intensity of the line are 
left free to vary, the best fit values (rest frame) are E = 6.35$^{+0.08}_{-0.03}$ keV 
(almost neutral) and EW = 165$\pm{49}$ eV. 
%
The upper limit to the line width is about 100 eV, which is 
consistent with the SIS energy resolution at the time the observation was carried out. 
%
%
The data to model ratio (Fig.~\ref{lab3}) clearly suggests an absorption feature above 7 keV. 
Since there is no instrumental level of uncertainty larger than 10\% around this energy and 
the background spectra are taken from source-free regions nearby IRAS~04575$-$7537, 
one can confidently assume that this feature is a real physical structure in the data. 
The addition of an iron edge at E = 7.13$^{+0.21}_{-0.16}$ keV 
results to be significant at $>$ 99.9\% level 
($\Delta\chi^{2}$ = 18 with two more degrees of freedom), 
according to an F--test. 
Iron K edges have been recently detected by ASCA in some other 
Seyfert 2 galaxies (Turner et al. 1997). The values of both the edge energy and 
absorption depth are similar to the present ones.\\
%
Regardless of the model used to fit the 0.5--10 keV spectrum, the photon index 
remains flat ($\Gamma$ = 1.5$\pm{0.05}$, Fig.~\ref{lab4}) and incompatible with the 
Seyfert 1s value of 1.9$\pm{0.1}$ (Nandra \& Pounds 1994; Nandra et al. 1997). 
The addition of a reflection component brings no improvement in 
the spectral fitting (Table 2, lines 4 and 5), even if the photon index is forced to be steep 
($\Gamma$ = 1.9). In a similar way, a warm absorber model applied to the 
hard component is not a satisfactory explanation for IRAS~04575$-$7537 
complex spectrum, the absorption edge being still highly significant. 
%
%

\section{Discussion}

\subsection{X-ray continuum properties}

The intrinsic hard X-ray luminosity of IRAS~04575$-$7537 
($L_{\rm X}$ $\simeq$ 2 $\times$ 10$^{43}$ \mbox {erg s$^{-1}$} in the 2--10 keV 
energy range) and the variability suggest the presence of a Seyfert 1 nucleus. 
The optical spectrum does not reveal 
any broad component, confirming that absorption effects play an important 
role. 
The best fit spectral slope, $\Gamma$ $\sim$ 1.5, is rather flat and similar to the 
value found in other flat-spectrum Seyfert 2 galaxies 
(Smith \& Done 1996; Cappi et al. 1996; Iwasawa et al. 1997), thus increasing the number of 
Seyfert 2 galaxies with flat spectra that are apparently inconsistent with the predictions 
of the Unified models. 
A simple extrapolation of the X-ray spectral slope up to 100 keV 
gives a 2--100 keV luminosity of about 6.0 $\times$ 10$^{43}$ 
\mbox {erg s$^{-1}$}, 
in excellent agreement with the far-infrared (40--120 $\mu$m) luminosity 
($L_{\rm FIR}$ $\simeq$ 6.2 $\times$ 10$^{43}$ \mbox {erg s$^{-1}$}, 
Boller et al. 1992). This seems to confirm the validity of a reprocessing scenario, 
in which most of the optical-UV-X--ray radiation emitted by the central source 
is intercepted by the torus, which reemits a fraction of this energy 
in the infrared band (Krolik \& Lepp 1989; Pier \& Krolik 1992). 

\subsection{Origin of the iron line and edge}

%
Assuming the iron cross section given by Leahy \& Creighton (1993), the measured 
absorption depth $\tau$ = 0.33$^{+0.14}_{-0.13}$ implies an equivalent hydrogen 
column density $N_{\rm H}$ = 2--4 $\times$ 10$^{23}$ cm$^{-2}$, which is much 
larger than the value obtained from the spectral analysis 
($N_{\rm H}$ $\sim$ 10$^{22}$ cm$^{-2}$). 
The lack of a reflection component seems to indicate that the iron line takes its origin 
through transmission, but 
both the observed {\rm Fe}\,K$\alpha$ edge depth and line EW 
are found to be inconsistent with the measured absorption column density 
unless an unrealistic factor $\sim$ 10 iron overabundance is assumed. 

\begin{table*}
\caption[]{``Leaky-total absorber'' model. 
N$_{{\rm H}_{1}}$ is the column density of the totally covering material, 
while N$_{{\rm H}_{2}}$ represents the absorption due to the partially covering one}
\begin{flushleft}
\begin{tabular}{ccccccc}
\noalign{\smallskip}
\hline
\noalign{\smallskip}
$\Gamma$&N$_{{\rm H}_{1}}$&$E_{\rm line}$&EW&N$_{{\rm H}_{2}}$
&Cvr Fract&$\chi^{2}/dof$\\
\noalign{\smallskip}
&(10$^{22}$ cm$^{-2}$)&(keV)&(eV)&(10$^{22}$ cm$^{-2}$)
&(\%)&\\
\noalign{\smallskip}
\hline \hline
\noalign{\smallskip}

1.9 (f)&1.22$^{+0.05}_{-0.06}$&6.35$^{+0.08}_{-0.03}$
&158$^{+47}_{-48}$&10.9$^{+4.20}_{-3.37}$&
36.1$^{+3.5}_{-3.8}$&832/811\\
%
%
\noalign{\smallskip}
\hline
\end{tabular}
\end{flushleft}
\end{table*}
The complex picture that derives from data analysis could be 
explained in a way similar to the model proposed by Hayashi et al. (1996) for the 
Narrow Emission Line Galaxy (NELG) NGC~2110 
and previously by Weaver et al. (1994) for NGC~4151. 
This requires the presence of a dual-absorber model, in which 
the iron features revealed by ASCA originate in a two-phase cold matter. 
Such a model has been successfully fitted to the IRAS~04575$-7537$ spectrum and the results are 
reported in Table 3. 
Cold material with a 
column density of the order of 10$^{23}$ cm$^{-2}$ partially covers the 
nucleus, with an average covering factor value of about 36 \%, and is responsible
for a substantial part of the iron line and edge features. An additional 
component with $N_{\rm H}$ $\sim$ 10$^{22}$ cm$^{-2}$, totally covering the central source, is 
responsible for the optical and soft X-ray absorption and provides a further contribution 
to the X-ray features. 
Since the intrinsic luminosity ($L_{\rm X}$ $\sim$ 2 $\times$ 10$^{43}$ \mbox 
{erg s$^{-1}$}) is typical of a Seyfert 1 nucleus, the underlying X-ray continuum 
is parameterized by a power law with spectral index 1.9 (Nandra \& Pounds 1994; Nandra et al. 1997). 
Following the calculations of Leahy \& Creighton (1993, see their Fig.~2) the 
measured absorbing column density predicts a {\rm Fe}\,K$\alpha$ line EW and 
edge depth consistent with the measured values, once they are scaled for the covering 
fraction obtained from the spectral fitting.\\
%
%
As a proof of this result, 
the addition of a {\rm Fe}\,K$\alpha$ edge is no longer required by the data, since  
a column density 
of the order of 10$^{23}$ cm$^{-2}$ can account for the iron edge.\\
Even if this model, that we call ``leaky-total absorber'', 
provides a greater $\chi^{2}$ (Table 3) than the previous absorbed 
power law plus iron edge fit (Table 2, line 3), 
it is interesting to note that it explains the iron features 
and offers a straightforward physical interpretation consistent with the predictions 
of Unified models as well. 
The partially ($\sim$ 36\%) covering material 
($N_{\rm H}$ $\sim$ 10$^{23}$ cm$^{-2}$) could indeed 
be associated either with the BLR clouds or with blobs, under the assumption that 
they are small if compared with both the size of the X-ray source and the radius of their 
distribution. This may be possible in a context of confinement by strong 
equipartion magnetic fields (Rees 1987) or by radiation pressure 
induced by the primary source (Celotti et al. 1992). 
Also the line width ($\sigma$ $\leq$ 100 eV, corresponding to $\sim$ a few 
$\times$ 10$^{3}$ 
Km s$^{-1}$) is consistent with typical values for BLR clouds 
(Ghisellini et al. 1994). 
Moreover the BLR are known from optical observations (Kwan \& Krolik 1981) 
to have column densities just of the order of those observed in the X-ray spectrum. 
The totally 
covering absorber could be associated with the obscuring torus that hides 
the active nucleus. 
The derived column density requires that either the torus is 
intrinsically optically thin 
or we are looking through its rim, 
which is consistent with the observed source variability. 
%

An alternative, but similarly speculative, explanation of IRAS~04575$-$7537 
geometry and X-ray emission introduces a more complex configuration 
for the torus (Maiolino \& Rieke 1995). 
If the gravitational potential is dominated by nuclear mass, 
as in the case of AGNs, 
then the Roche limit requires that the inner parts of the 
torus are tidally disrupted when they approach a distance (Lang 1980)
\begin{displaymath}
R\left(pc\right) = 2.5\left[M_{\mathrm {BH}}\left(M_{\odot}\right)/\rho
\left(cm^{-3}\right)\right]^{1/3}
\end{displaymath}
Thus, when a typical molecular cloud in a Seyfert galaxy moves within this radius, 
it will be destroyed by the nuclear black hole gravitational potential. 
This can explain the partially covering absorber, whereas the outer parts of 
the torus remain quite homogeneous and can be associated with 
the totally covering absorber. 

\subsection{The soft excess}

In order to model the soft X-ray emission,  
the 0.4--10 keV spectrum was fitted adding to the high-energy best-fit spectrum a soft component, 
either a thermal plasma or a power law model, and fixing the absorption of the soft component 
at the Galactic value ($N_{{\rm H}_{\rm gal}}$ = 9 $\times$ 10$^{20}$ cm$^{-2}$, 
Dickey \& Lockman 1990). 
%
%
Due to the low number of source counts below 1 keV, it has not been possible to 
discriminate between a thermal component or a scattering power law model 
for the soft X-ray excess from the present data. 
%
%
The 0.5--4.5 keV X-ray luminosity derived from the thermal model is 
$\sim$ 4.3 $\times$ 10$^{41}$ \mbox {erg s$^{-1}$} and is likely to be associated 
to hot gas in the host galaxy. 
This value results to be more than an order of magnitude larger than the one 
($L_{\rm X}$ $\sim$ 3.8 $\times$ 10$^{40}$ \mbox {erg s$^{-1}$}) expected 
for bright infrared normal and starburst galaxies, assuming the far-infrared/X-ray 
luminosity correlation of David et al. (1992). 
This suggests that IRAS~04575$-$7537 soft X-ray excess is due to 
scattered nuclear X-ray emission. 
In this context, a scattering model 
gives a rather good description of the data (Table 2, line 6), 
and provides a fraction of $\sim$ 3\% of scattered emission. 
%
%
%
Archival ROSAT PSPC data have also been analyzed. The source is about 18$\arcmin$ from 
the center of the field of view. The spectral analysis 
confirms that the spectrum is highly absorbed ($N_{\rm H}$ $\sim$ 10$^{22}$ cm$^{-2}$), 
but the existence of a soft excess appears marginal, 
since the high galactic column density 
hampers a spectral study below $\sim$ 0.7 keV. 
It should be noted that the high soft X-ray luminosity (about 10$^{43}$ \mbox{erg s$^{-1}$}) 
found by Moran et al. (1996) 
must be ascribed to the ROSAT 1--2 keV flux, where the hard luminous X-ray 
component is already significant. 

\section{Conclusions}

The ASCA observation of the Sey 2 galaxy IRAS~04575$-$7537 has revealed a complex spectrum. 
Relevant points may be summarized as follows:\\
\begin{enumerate}
\item Hard X-ray variability, of a factor $>$ 2 over a time scale of about 5 $\times$ 10$^{4}$ s, 
has been detected, without a significant evidence of spectral variability. 

\item The hard (E $>$ 1 keV) X-ray continuum is well described by an absorbed 
power law model, whose flat spectral slope ($\Gamma$ $\sim$ 1.5) cannot be accounted for 
by a strong reflection component on a steeper primary power law, suggesting a possible 
inconsistency with the predictions of the Unified models. 

%
\item 
Iron K$\alpha$ absorption edge and emission line have been clearly detected. 
Both of them are consistent with neutral or mildly ionized (less 
than \ion{Fe}{xvii}) iron, but are too strong to be explained by transmission 
through the observed absorption column density ($N_{\rm H}$ $\sim$ 10$^{22}$ cm$^{-2}$). 



\item Excess X-ray emission below 1 keV is detected, but the present data do not allow to 
discriminate between a thermal and a scattering model. 

\item 
The whole observation can be explained assuming a model, called ``leaky-total absorber'', 
which consists of a dual-absorber configuration, 
where the underlying continuum is parameterized to be that of a Seyfert 1 ($\Gamma$ = 1.9). 
A partially ($\sim$ 36\%) 
covering material, either BLR clouds or the inner, disrupted parts of the torus, 
surrounds the source and gives rise to an absorption of about 10$^{23}$ 
cm$^{-2}$. This material is considered to be responsible for the iron line and edge 
features. Moreover, the totally covering matter, probably associated with the molecular 
torus (or the outer parts of it), produces the optical and the soft X-ray absorption 
($N_{\rm H}$ $\sim$ 10$^{22}$ cm$^{-2}$). 
This model explains the observed spectral features, has the advantage to have a straightforward 
physical interpretation in the framework of Unified models and provides a solution of the 
apparent puzzle of points 2 and 3. 
%

\end{enumerate}

\begin{acknowledgements}
We thank all the members of the ASCA team who operate the satellite and 
maintain the software and database. 
This work has made use of the NASA/IPAC Extragalactic Database (NED)
which is operated by the Jet Propulsion Laboratory, Caltech, under contract
with the National Aereonautics and Space Administration, of data obtained throu gh the 
High Energy Astrophysics Science Archive Research Center Online Service, provided 
by the Goddard Space Flight Center and of the Simbad database, operated at CDS, Strasbourg, 
France. 
Financial support from Italian Space Agency under the contract ARS--96--70 
and MURST is acknowledged by CV, AC, MC and GGCP. 
The authors are grateful to R. Pazzaglia for the helpful 
suggestions about the ``leaky-total absorber'' model. 
Moreover, CV wishes to thank all those people who encouraged him to write this 
paper.
\end{acknowledgements}

\end{document}